\documentclass[aps,pre,twocolumn,showpacs,groupedaddress]{revtex4}

\usepackage{graphicx}
\usepackage{dcolumn}
\usepackage{bm}
\usepackage{amssymb}

\newcommand{\ie}{\mbox{\it i.\,e.}~}

\begin{document}


\title{Relevance of angular momentum conservation in mesoscale hydrodynamics simulations}

\author{Ingo O. G\"{o}tze}
\email[]{e-mail: i.goetze@fz-juelich.de}
\author{Hiroshi Noguchi}
\author{Gerhard Gompper}
\affiliation{
Institut f\"ur Festk\"orperforschung, Forschungszentrum J\"ulich, 
52425 J\"ulich, Germany}

\date{Received: date / Revised version: date}

\begin{abstract}
The angular momentum is conserved in fluids with a few exceptions such as ferrofluids.
However it can be violated locally in fluid simulations to reduce computational costs.
The effects of this violation are investigated using a
particle-based simulation method, multi-particle collision dynamics, 
which can switch on or off angular-momentum conservation.
To  this end, we study circular Couette flows between concentric and eccentric cylinders,
where non-physical torques due to the lack of the angular-momentum conservation 
are found whereas the velocity field is not affected.
In addition, in simulations of fluids with different viscosities in contact and star polymers 
in solvent, incorrect angular velocities occur.
These results quantitatively agree with the theoretical predictions based on the macroscopic stress tensor.
\end{abstract}


\maketitle

\section{Introduction}

In simulations of the hydrodynamic behavior of complex fluids, one is faced with the challenge of
bridging the gap between the mesoscopic length and time scales of the solute and the atomic scales
of the solvent.
As these length scales typically differ by orders of magnitude, a full treatment on a microscopic level
is prohibited by the huge number of involved particles and the large necessary time range.
Moreover, one is often only interested in the dynamics of the colloidal particles,
while the microscopic details of the solvent that mediates
the hydrodynamic interactions are rather unimportant.
Thus, a coarse-grained mesoscopic fluid model is required that is sufficiently simple
to be tractable but still captures the correct hydrodynamic behavior.

Various mesoscopic approaches have been proposed in the last decades.
A large number of physical solvent molecules is represented by one model fluid particle at a time,
reducing the number of degrees of freedoms considerably.
Lattice methods, such as lattice gas automata (LGA)~\cite{fris86} and
lattice-Boltzmann methods (LB)~\cite{succ01,yeom06}, generally suffer
from the lack of Galilean invariance.
Moreover, it is difficult to incorporate complex and deformable boundaries that play important roles in
the phase separation of two fluids~\cite{yeom06,tana00}
and the dynamics of vesicles and cells~\cite{skal90}.
In particle-based techniques such as
dissipative particle dynamics (DPD)~\cite{groo97,espa98,nogu07a} or
multi-particle collision dynamics (MPC)~\cite{yeom06,male99,ihle01,lamu01,alla02,kiku03,ihle03b,ihle05,padd04,padd06,hech05,male00b,ripo04,ripo06,webs05,lee06,wata07,nogu05b,nogu06a,saka02,tucc04,nogu07},
particle positions and velocities are continuous variables that are updated at discrete times.
Coupling to solute particles as well as moving boundaries can be easily treated.
MPC needs less computational time compared to other particle based methods such as DPD,
thus allowing simulations of larger systems.

In this article, we will focus on MPC,
which has been applied to various systems such as colloids~\cite{padd04,padd06,hech05}, polymers
~\cite{yeom06,male00b,ripo04,ripo06,webs05,lee06,wata07}, membranes~\cite{nogu05b,nogu06a},
ternary amphiphilic fluids~\cite{saka02}, and chemical reaction systems~\cite{tucc04}.
The MPC method naturally contains thermal fluctuations.
Hybrid simulations combining a MPC fluid with molecular dynamics (MD) of solute particles are
easily possible. The algorithm is constructed in such way that mass, energy and translational momentum
are locally conserved, which is essential for correct hydrodynamic behavior.
However, the {\it angular} momentum is not conserved in the most widespread version of MPC,
which is often called stochastic-rotation dynamics (SRD).
Here we denote it as MPC-SR.
The consequences of this fact have not yet been investigated and are the subject of this paper.
In order to clarify the effects of angular-momentum conservation, we mainly use the
Andersen-thermostat version of MPC, denoted MPC-AT, where angular momentum conserving
and non-conserving algorithms are available~\cite{nogu07}.
We also checked that the same quantitative dependence appears in the original MPC-SR method.
The main conclusion is, that simulations that do not conserve angular momentum can lead to quantitatively and even qualitatively incorrect results,
when the boundary conditions on walls are given by forces, fluids with different viscosities are in contact,
or finite-sized objects rotate in fluids.

The rest of this paper is organized as follows:
In Sec. \ref{sec:fluid_dyn}, we briefly discuss
the effect of the non-conservation of angular momentum on the stress tensor.
Note, that while in the MPC fluid the non-conservation of angular momentum is
an artifact of the simulation method, there are also real fluids,
where angular momentum is not conserved.
For example, in ferrofluids asymmetric stress arises~\cite{rose85,kuik95} when the rotation of
the suspended particles is impeded by external fields.
In Sec.~\ref{sec:sim_met}, the algorithms for the angular-momentum conserving and non-conserving
versions of MPC-AT are described.
A simple geometry to study rotating fluids is the flow between rotating coaxial cylinders,
also called circular Couette flow.
The simulation results for the angular-momentum conserving and non-conserving methods
are compared in Sec.~\ref{sec:couette}.
In particular, binary fluid and branched polymer systems are investigated in Sec.~\ref{sec:binfl} and D, respectively.
In Sec.~\ref{sec:ecc}, we lift the restriction of coaxiality and study rotating eccentric cylinders.
This geometry is of practical importance in journal bearings and microfluidic devices using rotating colloids \cite{bech06}.
Finally, we summarize our results in Sec.~\ref{sec:sum}.

\section{Macroscopic fluid dynamics \label{sec:fluid_dyn}}

In conventional viscous fluids that do conserve angular momentum, the viscous stress tensor 
has to be symmetric, \ie $\sigma_{\alpha\beta}=\sigma_{\beta\alpha}$. 
This symmetry is required by the fact that there is no stress expected in a uniformly rotating fluid 
(rigid body rotation)~\cite{land87}, or alternatively, by the conservation 
of angular momentum~\cite{batc67}.
On the other hand, for a fluid without conservation of angular momentum, 
the above argument is no longer valid and we have to consider in general an asymmetric tensor.

Here, we consider Newtonian fluids, \ie the stress is proportional to the strain rate, 
so that the $\sigma_{\alpha\beta}$ are linear functions of the derivatives 
$\partial v_{\alpha}/\partial x_{\beta}$~\cite{land87}.
We decompose the stress tensor in its symmetric and asymmetric parts.
Then, the viscous stress is given by
\begin{eqnarray} \label{eq:strs}
\sigma_{\alpha\beta}&=&\lambda(\nabla \cdot {\bf v})\delta_{\alpha\beta} \\
&+& \bar{\eta} \left(\frac{\partial v_{\alpha}}{\partial x_{\beta}} \nonumber
+\frac{\partial v_{\beta}}{\partial x_{\alpha}} \right) + \check{\eta} \left(\frac{\partial v_{\alpha}}{\partial x_{\beta}}-\frac{\partial v_{\beta}}{\partial x_{\alpha}} \right),
\end{eqnarray}
where $\alpha,\beta \in \{x,y,z\}$. Here, 
$\lambda$ is the second viscosity coefficient, and $\bar{\eta}$ and $\check{\eta}$ 
are the symmetric and asymmetric components of the viscosity, respectively.
The last term in Eq.~(\ref{eq:strs}) is linear in the vorticity 
$\nabla\times{\bf v}$,
and does not conserve angular momentum.
Thus, the last term vanishes (\ie $\check{\eta}=0$) in angular-momentum-conserving systems.

The equation of velocity evolution is given by
\begin{equation} \label{eq:nseq}
\rho\frac{D {\bf v}}{D t}= -\nabla P + (\lambda+ \bar{\eta}-\check{\eta})\nabla(\nabla\cdot{\bf v})
+ (\bar{\eta}+\check{\eta})\nabla^2 {\bf v},
\end{equation}
where $D/Dt$ is Lagrange's derivative and $P$ is the pressure.
When a fluid is incompressible,
this is the normal Navier-Stokes equation with viscosity $\eta=\bar{\eta}+\check{\eta}$.
This is consistent with the usual definition of the shear viscosity  $\eta=\sigma_{xy}/\dot\gamma$ 
in simple shear flow with the velocity field ${\bf v}=\dot\gamma y {\bf e}_x$.
Since the equations of continuity and velocity evolution are of the same form,
the negligence of angular-momentum conservation does not modify the velocity field of fluids
when the boundary conditions are given by velocities.
However, it generates an additional torque, 
so that the velocity field can be changed when the boundary condition is given by forces.
In cylindrical coordinates ($r,\theta,z$), the azimuthal stress is given by
\begin{equation} \label{eq:strs_rq}
\sigma_{r\theta}=(\bar{\eta}+\check{\eta})
\frac{r\partial (v_{\theta}/r)}{\partial r}
+ 2\check{\eta}\frac{v_{\theta}}{r}.
\end{equation}
The first term is the stress of the  angular-momentum-conserving fluid,
which depends on the derivative of the angular velocity $\Omega=v_{\theta}/r$.
The second term is
the additional stress from the negligence of angular-momentum conservation
and is proportional to $\Omega$.

When a fluid is compressible and the fluid density is not constant,
the bulk viscosity is not negligible.
The bulk viscosity without angular momentum conservation is given by
$\lambda+ 2\bar{\eta}/3$ instead of $\lambda+ 2\eta/3$.
In angular-momentum-conserving fluids, these two values coincide because $\eta=\bar{\eta}$.
Thus, the effects of the angular-momentum conservation are not negligible
when the torque on objects or the bulk viscosity is significant in fluid systems.
Eqs.~(\ref{eq:strs}--\ref{eq:strs_rq}) are general and can be applied to MPC methods and other model fluids, 
which do not conserve angular momentum.
We explain the effects of the torque quantitatively using MPC-AT in the following sections.

\section{Simulation Method \label{sec:sim_met}}

\subsection{Multi-Particle Collision Dynamics}

MPC is one of the particle-based methods to simulate hydrodynamic behavior accompanied by thermal fluctuations.
A fluid is described by point-like particles of mass $m$.
The MPC algorithm consists of alternating streaming and collision steps. 
In the streaming step, the particles move ballistically, 
${\bf r}_{i}(t+\Delta t) 
    = {\bf r}_{i}(t) + {\bf v}_{i} \Delta t$,
where $\Delta t$ 
is the time interval between collisions.
Subsequently, the particles are sorted into the cells of a cubic lattice with lattice constant $a$ 
that is randomly shifted before each collision step to ensure  
Galilean invariance~\cite{ihle01}.
The collision step then mimics the simultaneous interaction of all particles within each cell 
by assigning the particles new velocities.
There are several versions of the collision procedures and 
each version can switch on or off angular-momentum conservation~\cite{nogu07,nogu07b}.
We call the versions of methods with or without angular-momentum conservation '$+a$' or '$-a$', respectively.
In the original version (MPC-SR), the relative particle velocities with respect to 
the mean velocity in a cell are rotated by a 
fixed angle $\varphi$ around an axis, which is chosen randomly for each 
cell~\cite{male99}. 
In MPC-AT$-a$, the velocities of the particles are updated by~\cite{alla02,nogu07}
\begin{equation}\label{eq:mpcat}
{\bf v}_{i}^{\rm {new}}= {\bf v}_{\rm c}^{\rm G} + {\bf v}_{i}^{\rm {ran}} 
  - \sum_{j \in {\rm cell}} {\bf v}_j^{\rm {ran}}/N_{\rm {c}},
\end{equation}
where $N_{\rm {c}}$ is the number of particles in a cell, and
velocities ${\bf v}_{i}^{\rm {ran}}$ are chosen from a 
Maxwell-Boltzmann distribution.
The center-of-mass velocity ${\bf v}_{\rm c}^{\rm G}$ of each cell is conserved, and
the temperature is constant in MPC-AT.
In MPC-AT$+a$, the velocities of the particles are updated by~\cite{nogu07}
\begin{eqnarray}\label{eq:mpcatan}
{\bf v}_{i}^{\rm {new}} &=& 
 {\bf v}_{\rm c}^{\rm G} + {\bf v}_{i}^{\rm {ran}}
 - \sum_{j \in {\rm cell}} {\bf v}_j^{\rm {ran}}/N_{\rm {c}} \\ \nonumber
&& +  m{\bf \Pi}^{-1} \sum_{j \in {\rm cell}} \left\{ {\bf r}_{j,{\rm c}}\times 
     ({\bf v}_j-{\bf v}_j^{\rm {ran}})\right\}\times {\bf r}_{i,{\rm c}},
\end{eqnarray}
where ${\bf \Pi}$ is the moment-of-inertia tensor of the particles in the cell.
The relative position is ${\bf r}_{i,{\rm c}}={\bf r}_i - {\bf r}_{\rm c}^{\rm G}$
where ${\bf r}_{\rm c}^{\rm G}$ is the center-of-mass of all particles in the cell.

\subsection{Boundary Conditions \label{sec:boundary}}
In order to simulate no-slip boundary conditions,
the following technique has been developed for $-a$ fluids in Ref.~\cite{lamu01}.
In the streaming step,
the fluid particles are scattered with a bounce-back rule on surfaces.
In the collision step, in collision cells crossing a boundary with $N_{\rm {c}}<n=\langle N_{\rm c}\rangle$,
a virtual particle with mass $m(n-N_{\rm {c}})$ and velocity 
${\bf v}_{\rm {wall}}+{\bf v}^{\rm {ran}}/(n-N_{\rm {c}})$ 
is inserted to calculate ${\bf v}_{\rm c}^{\rm G}$, where ${\bf v}_{\rm {wall}}$ is the
velocity of the boundary wall.
This algorithm keeps the slip on a boundary small~\cite{lamu01}.

We have tested some algorithms for $+a$ methods,
where the position of the virtual particle is now important.
One possibility (denoted `cen') is to locate it at the center of the cell.
For a simple geometry like a cylinder, 
more sophisticated ways are available, \mbox{\it e.\,g.}~by putting a virtual particle slightly inside boundary walls,
which can reduce slip.
A more direct way to estimate the velocities inside a wall
is to distribute explicit particles inside the wall.
Watari {\it et\ al.}~\cite{wata07} proposed a boundary algorithm,
where particles freely enter inside objects and velocities of inside particles 
are updated to ${\bf v}^{\rm {ran}}$.
However, this allows flows to penetrate through a small object,
when there is a pressure difference around the object.
To prevent flow penetration, we  employ the bounce-back rule.
Particles are randomly distributed inside the cylinder wall with depth $\sqrt{2}a$ from the surface
 with the same density as the outside fluid.
Before collision steps, the velocity is updated to ${\bf v}_{\rm {wall}}+{\bf v}^{\rm {ran}}$.
The position of the wall particles are updated by renewal of the random uniform distribution
\cite{footnote1}.
In the Couette flow simulations, the velocity field is theoretically known.
Thus, ${\bf v}_{\rm {wall}}$ is extrapolated for a wall-particle position in most of the simulations 
(denoted '$\omega$-gra').
This explicit-particle boundary algorithm can be applied to other particle-based methods such as DPD.
We employ '$\omega$-gra' and `cen' algorithms for coaxial systems (Sec.~\ref{sec:couette}) 
and eccentric cylinders (Sec.~\ref{sec:ecc}), respectively.
We show the comparison of these two boundary algorithms and '$\omega$-con' algorithm for Couette flow in Sec.~\ref{sec:2rot}.
In '$\omega$-con', explicit wall particles with
the constant angular velocity $\Omega_{\rm {bd}}$ are employed so that
 ${\bf v}_{\rm {wall}}=\Omega_{\rm {bd}} r_i {\bf e}_{\theta}$.

\begin{figure}
\includegraphics{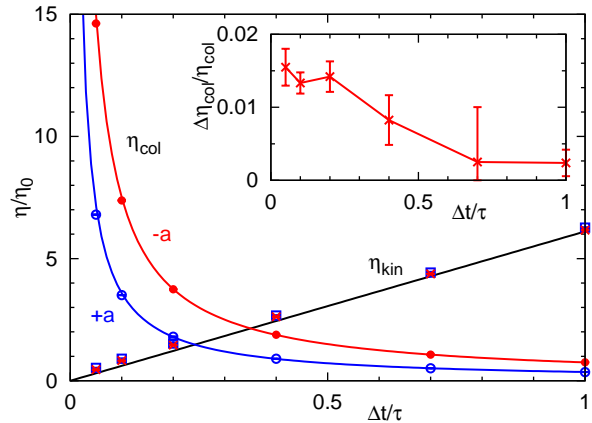}
\caption{\label{fig:vis}
(Color online)
Dependence of the viscosity $\eta$ of MPC-AT$-a$ ($\bullet$, $\times$) and MPC-AT$+a$ ($\circ$, $\Box$) on the time step $\Delta t$ for $n=10$ in two dimensions.
The inset shows the viscosity difference $\Delta \eta_{\rm {col}}=\bar{\eta}_{\rm {col}}-\check{\eta}_{\rm {col}}$
of MPC-AT$-a$.
}
\end{figure}

\subsection{Viscosity}
The shear viscosity is calculated from $\sigma_{xy}/\dot\gamma=\eta=\bar{\eta}+\check{\eta}$ 
in simple shear flow with ${\bf v}=\dot\gamma y {\bf e}_x$.
The viscosity of MPC consists of two contributions, $\eta=\eta_{\rm {kin}}+\eta_{\rm {col}}$;
the kinetic viscosity $\eta_{\rm {kin}}$ and the 
collision viscosity $\eta_{\rm {col}}$ result from the momentum transfer
due to particle displacements and collisions, respectively.
The viscosity of MPC-AT$-a$ with large mean number density $n$,
is given by~\cite{nogu07}
\begin{eqnarray} 
\eta_{\rm {kin}} &=& \frac{nk_{\rm B}\Theta\Delta t}{a^d}\left(\frac{n}{n-1}-\frac{1}{2} \right) \\
\eta_{\rm {col}} &=& \frac{m(n-1)}{12a^{d-2}\Delta t}
\end{eqnarray}
where $d$ and $k_{\rm B}\Theta$ are the spatial dimension,
and thermal energy, respectively.
The viscosity of MPC-AT$+a$ with large $n$ can be calculated similarly, and is found to be
\begin{eqnarray} 
\eta_{\rm {kin}} &=& \frac{n k_{\rm B}\Theta \Delta t}{ a^d }
\left[ \frac{n}{n- (d+2)/4 }
          - \frac{1}{2} \right], \\
\eta_{\rm {col}} &=& \frac{m(n-7/5)}{24 a^{d-2}\Delta t}.
\end{eqnarray}
The derivation and the correction terms for small $n$ for $+a$ versions of the MPC family
will be reported elsewhere~\cite{nogu07b}.
In two-dimensional systems, the angular-momentum constraint does not change 
the kinetic viscosity for large $n$, \ie $\eta_{\rm {kin}}^{\rm {+a}}\simeq \eta_{\rm {kin}}^{\rm {-a}}$.
We also calculate the viscosity from simulations for simple shear flow
with Lees-Edwards boundary conditions~\cite{alle87}.
Fig.~{\ref{fig:vis}} shows that the theoretical and 
numerical results are in very good agreement.

The symmetric and asymmetric components of shear viscosity $\bar{\eta}$ and $\check{\eta}$
are calculated from the shear stress $\sigma_{yx}/\dot\gamma=\bar{\eta}-\check{\eta}$.
Since the kinetic stress is symmetric in $x$ and $y$, \ie $\sigma_{yx}^{\rm {kin}}=\sigma_{xy}^{\rm {kin}}$, 
the kinetic viscosity has no asymmetric component $\check{\eta}_{\rm {kin}}=0$.
The collision procedure of MPC-AT$-a$ does not conserve the angular momentum.
The molecular chaos assumption gives
$\sigma_{yx}^{\rm {col}}=0$, because $\langle v_y(x)\rangle=0$ before and after the collisions.
Thus, the viscosities are 
\begin{equation}
\check{\eta}=\bar{\eta}_{\rm {col}}=\eta_{\rm {col}}/2. 
\end{equation}
This viscosity relation holds for all $-a$ versions of MPC and DPD in Refs.~\cite{nogu07,nogu07b}.
The numerical simulation of MPC-AT$-a$ shows good agreement with a deviation of only about $1$\% for $n=10$
(see the inset of Fig.~\ref{fig:vis}).

\subsection{Parameters}
We simulate two-dimensional flows.
The simulation data are displayed with the units of length $a$, time $\tau=a\sqrt{m_0/k_{\rm B}\Theta}$, and viscosity $\eta_0=\sqrt{m_0k_{\rm B}\Theta}/a$.
We use $n=10$ and $\Delta t/\tau=0.05$ or $0.1$.
Since our aim is to clarify the difference of viscous stresses between $+a$ and $-a$ fluids,
we use small angular velocities $\Omega\tau= 0.004$ to $0.01$
for circular Couette flows
to keep the density constant and a low Reynolds number $Re=\rho D^2\Omega/\eta\approx 1$,
where $D=10a$ is the diameter of the smaller cylinder.
To obtain the hydrodynamics of liquids,
we use a small Knudsen number $Kn=l_{\lambda}/D=0.01$,
where $l_{\lambda}=\Delta t\sqrt{k_{\rm B}\Theta/m_0}$ is the mean free path of fluid particles.
The error bars are estimated from three or ten independent runs.

MPC-AT is more time consuming than
MPC-SR due to the heavier use of random numbers ($dN_{\rm c}$ Gaussian-distributed instead of $d-1$ uniformly-distributed random numbers).
On the other hand, taking angular-momentum conservation into account only slightly increases the required CPU time 
in two dimensional simulations.

\begin{figure}
\includegraphics{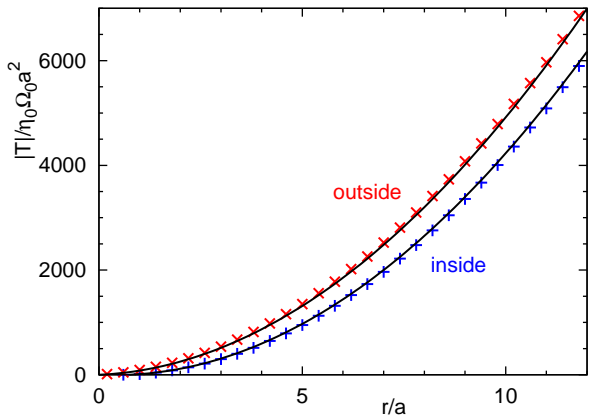}
\caption{\label{fig:torq1}
(Color online)
Torque $T$ in a rotating MPC-AT$-a$ fluid with uniform angular velocity $\Omega_0$ for $n=10$ and $\Delta t=0.1$.
Symbols $+$ and $\times$ represent the torque on the inner and outer surfaces of a virtual cylinder in the fluid, respectively.
Solid lines are obtained by Eq.~(\ref{eq:Tfin}).
The torques for the inner and outer surfaces have opposite signs; for clarity, only the absolute values are shown.
Error bars are smaller than the size of the symbols.
}
\end{figure}

\section{Circular Couette Flow \label{sec:couette}}
We consider Couette flow, since it is a well analyzed, simple system.
Let $R_1$ and $R_2$ be the radii of two coaxial cylinders rotating with the angular frequencies $\Omega_1$ and $\Omega_2$ respectively, 
where the indices 1 and 2 refer to the inner and outer cylinders, respectively.
We assume both cylinders to be of infinite length and their angular velocities to be sufficiently low, such that no Taylor-Couette instabilities occur, 
and the problem can be considered in two dimensions.
For symmetry reasons, the radial velocity component vanishes and 
the Navier-Stokes equation yields the azimuthal velocity \cite{trit88}
\begin{equation} \label{eq:v_cf}
v_\theta(r) = Ar+ B/r
\end{equation}
where
\begin{equation}
A = \frac{\Omega_2 R_2^2 - \Omega_1 R_1^2}{R_2^2 - R_1^2} {\rm \ and\ }
B = \frac{(\Omega_1 - \Omega_2) R_1^2R_2^2}{R_2^2 - R_1^2}.
 \label{eq:v_cfab}
\end{equation}
The torques acting on the cylinders in an $+a$ fluid, which conserves angular momentum are \cite{trit88}
\begin{equation}
T_1  = - T_2 =  \frac{4 \pi \mu R_1^2 R_2^2 (\Omega_2 - \Omega_1)}{R_2^2 - R_1^2} \label{T1}\\
 \label{eq:T1-T2}
\end{equation}
The torque $T$ in a fluid at radius $R_1<r<R_2$ is calculated from the momentum transfer across a virtual cylinder of radius $R=r$,
and is equal to $T_1$ or $-T_1$ on the inner or outer surface of the virtual cylinder, respectively.
Thus, the torque on the inner cylinder propagates to the outer cylinder via the fluid with a constant value 
because of angular-momentum conservation.
However, in an $-a$ fluid, the negligence of the angular-momentum conservation generates an additional torque.

\subsection{Uniform Angular Velocity}
First, we consider the simplest case, where the whole fluid rotates with constant angular velocity $\Omega_0$.
This is done with $R_1=0$ and $\Omega_2\equiv\Omega_0$ or 
both cylinders rotate with the same angular velocity $\Omega_1=\Omega_2\equiv\Omega_0$.
Here, no torque is expected to be acting on the cylinders in $+a$ fluids, as this corresponds to the rotation of a rigid body.
The MPC-AT$+a$ simulations yield the physically correct result, $T=0$, at any $r$.
However, in the MPC-AT$-a$ and MPC-SR simulations, we {\it do} observe positive or negative torques on
the confining inner ($R=R_1$) or outer ($R=R_2$) cylinder, respectively.
In the following, we consider the torques on the inner and outer surfaces of a virtual cylinder of radius $R$ in the fluid,
which shows the torque generation in the $-a$ fluid.
In the MPC simulations, we calculate the torques on the inner and outer surface of this virtual cylinder by
measuring the change of the angular momentum per time step in cells crossing the virtual cylinder at $R=r$.
The results are shown in Fig.~\ref{fig:torq1}.
This torque is explained by the stress term of the asymmetric viscosity $\check{\eta}$ in Eq.~(\ref{eq:strs_rq}).
The torque is the tangential stress $2\check{\eta}\Omega_0$ multiplied by the circumference length $2\pi R$ and the radius $R$, \ie  
$|T|=4\pi\check{\eta}\Omega_0 R^2$.
The torque $T_{\rm {m}}=(|T_{\rm {in}}|+|T_{\rm {out}}|)/2$ averaged on inner and outer surfaces agrees with this prediction.

However, inner and outer surfaces of the cylinder receive slightly smaller and larger torques than $T_{\rm m}$.
This mismatch is qualitatively explained as follows.
The total transferred momentum of particles inside the cylinder is equal to that of outer particles
with the opposite sign,
since the translational momentum is conserved.
Thus, the torque of inner particles is smaller than the outer one,
since the average distance from the cylinder axis of inner particles is smaller.
In order to calculate this finite-cell-size effect quantitatively, 
we consider the transfer of momentum crossing a cylinder of radius $R$ in the fluid.
It is derived in analogy to the momentum crossing a plane in calculations of the viscosity~\cite{kiku03,ihle03b,ihle05,nogu07b},
and the details are described in the Appendix.
The resulting torques $T_{\rm {in}}$ and $T_{\rm {out}}$ of a virtual cylinder of radius $R$ 
that are exerted on the inner and outer surfaces, respectively, are found to be
\begin{equation}
T_{{\rm in},{\rm out}}(R) = \pm 4 \pi \check{\eta} \Omega R^2 \bigg(1 \mp \frac{3a}{4R} \bigg). \label{eq:Tfin}
\end{equation}
Thus, the first-order correction term is $\mp 3a/4R$.
The same correction term can also be derived for $-a$ versions of the other MPC methods.
This correction term well describes the torque difference between inner and outer surfaces (see Fig.~\ref{fig:torq1}).

\begin{figure}
\includegraphics{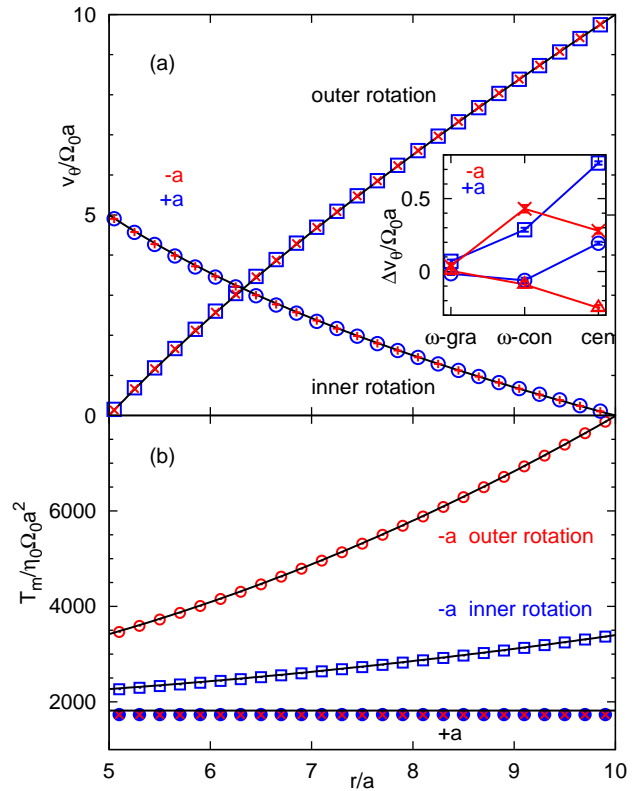}
\caption{\label{fig:v2rot}
(Color online)
(a) Azimuthal velocity $v_{\theta}$ and (b) averaged torque $T_{\rm {m}}=|T_{\rm {in}}|+|T_{\rm {out}}|)/2$ 
of circular Couette flow for $n=10$ and $\Delta t=0.1$.
The inner ($R_1=5a$) or outer ($R_2=10a$) cylinder rotates with $\Omega_0\tau=0.01$,
the other cylinder is fixed ($\Omega=0$).
(a) Symbols represent MPC-AT$+a$ ($\circ$, $\square$) and MPC-AT$-a$ ($+$, $\times$).
(b) Symbols represent MPC-AT$+a$ ($\bullet$, $\times$) and MPC-AT$-a$ ($\circ$, $\square$).
Solid lines are obtained by (a) Eqs.~(\ref{eq:v_cf}), (\ref{eq:v_cfab}) and (b) Eqs.~(\ref{eq:T1-T2}), (\ref{eq:strs_rq}).
Error bars are smaller than the size of the symbols.
The inset shows the slip velocity $\Delta v_{\theta}$ ($+a$, $-a$) on inner ($\Box$, $\times$) and outer ($\circ$, $\triangle$) cylinders 
in the outer cylinder rotation for the different boundary algorithms introduced in Sec. \ref{sec:boundary}.
}
\end{figure}

\subsection{Angular Velocity Gradient}\label{sec:2rot}
Next, we consider the flow with angular velocity gradient induced by 
$(\Omega_1,\Omega_2)=(0,\Omega_0)$ or $(\Omega_0,0)$.
Both $+a$ and $-a$ fluids yield the velocity field described by Eqs.~(\ref{eq:v_cf}) and (\ref{eq:v_cfab}) [see Fig.~\ref{fig:v2rot}(a)].
The torque $T$ in the $+a$ fluid is constant throughout the fluid, and
 depends only on the relative angular velocity $\Omega_1-\Omega_2$, not on the absolute value of $\Omega_1$ or $\Omega_2$. 
This results agrees with the prediction of Eq.~(\ref{eq:T1-T2}).
However, the torque $T$ in the $-a$ fluid is not constant and 
depends on the value of the angular velocity because of the non-conservation of angular momentum.
This dependence is well described by Eq.~(\ref{eq:strs_rq}) [see Fig.~\ref{fig:v2rot}(b)].

The inset of Fig.~\ref{fig:v2rot}(a) shows the slip velocity 
$\Delta v_{\theta}=v_{\theta}^{\rm {lsf}} - v_{\theta}^{\rm {th}}$ on the boundaries, 
where $v_{\theta}^{\rm {th}}$ is given 
by Eqs.~(\ref{eq:v_cf}) and (\ref{eq:v_cfab}).
The velocity $v_{\theta}^{\rm {lsf}}$ is calculated from a least-squares fit 
to Eq.~(\ref{eq:v_cf}) with parameters $A$ and $B$ for the range $6<r/a<9$.
The '$\omega$-gra' algorithm shows very small slip and the velocity in Fig.~\ref{fig:v2rot}(a) 
coincides with the theoretical values very well.
The '$\omega$-con' and 'cen' algorithms show larger slip and $\pm a$ fluids show similar dependence.

\begin{figure}
\includegraphics{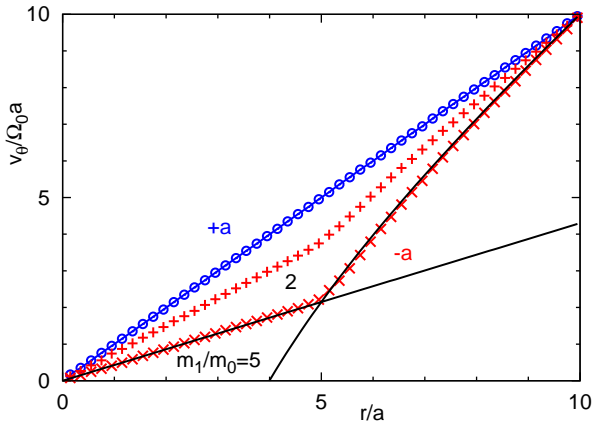}
\caption{\label{fig:2vis}
(Color online)
Azimuthal velocity of binary fluids in a rotating cylinder with $\Omega_0=0.01/\tau$.
The viscous fluids with particle mass $m_1$ and $m_0$ are located at $r<5a$ and $5a<r<10a$, respectively.
Symbols represent the simulation results of MPC-AT$-a$ with $m_1/m_0=2$ ($+$) or $m_1/m_0=5$ ($\times$), 
and MPC-AT$+a$ for $m_1/m_0=5$ ($\circ$).
Solid lines represent the analytical results for MPC-AT$-a$ at $m_1/m_0=5$.
Error bars are smaller than the size of the symbols.
}
\end{figure}

\subsection{Phase-Separated Binary Fluids}\label{sec:binfl}
A boundary of a fluid exists not only on solid objects but also between two fluids or on membranes.
In order to investigate the fluid-fluid boundary in $-a$ fluids,
we consider binary fluids with a fixed geometry of the boundary surface, which is impenetrable to the fluid particles.
The inner cylinder of radius $R_1$ of circular Couette flow is replaced by a more viscous fluid,
and the outer cylinder with radius $R_2$ rotates with constant velocity $\Omega_2=\Omega_0$.
This is a simplified description of oil and water phase-separated due to surface tension, 
or two liquids separated by a membrane.
It is assumed that cylinders rotate very slowly, and that the flow stress does not change the shape of the interface.
In MPC-AT, the fluids inside ($r<R_1=5a$) and outside ($R_1<r<R_2=10a$) have 
high viscosity $\eta_1$ with mass $m_1$ and low viscosity $\eta_2$ with mass $m_0$, respectively
 (note that $\eta\propto m$).
The particles of both fluids are scattered elastically at the boundary surface at $R_1$ during the streaming step,
but the MPC collision performed in cells crossing the boundary
propagates the momentum from one fluid to the other.

In MPC-AT$+a$, both fluids rotate with $\Omega_0$ independent of their viscosities.
However, in MPC-AT$-a$, the inner fluid rotates more slowly for $m_1 > m_0$ (see Fig.~\ref{fig:2vis}).
This is caused by the asymmetric stress term $2\check{\eta}\Omega$ for $-a$ fluids 
where $\check{\eta} \simeq \eta_{\rm {col}}/2$.
If both fluids rotate at the same angular velocity,
the inner and outer stresses do not coincide.
Thus, the angular velocity of the inner fluid $\Omega_1$ is smaller than the outer one.
The inner and outer flows are described by $v_\theta(r)=\Omega_1 r$ and
and Eq.~(\ref{eq:v_cf}), respectively.
Then, $\Omega_1$ is obtained from
the stress balance at $r=R_1$, \ie
$2\check{\eta}_1\Omega_1= (8/3)\eta_2 (\Omega_0-\Omega_1) +2\check{\eta}_2\Omega_1$.
This calculation well reproduces the numerical results  (see Fig.~\ref{fig:2vis}).
Thus, it is essential to employ an $+a$ version of MPC in simulations of multi-phase flows of binary fluids with different viscosities.

\begin{figure}
\includegraphics{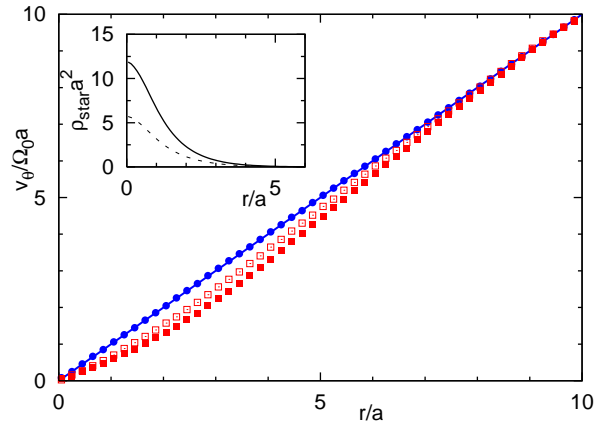}
\caption{\label{fig:star}
(Color online)
Azimuthal velocity $v_{\theta}$ of a fluid with a star polymer fixed in the center of a rotating cylinder.
Circles represent the simulation results for MPC-AT$+a$,
whereas the results for MPC-AT$-a$ are shown as squares, with full squares for $f=10$
and open squares for $f=5$.
The line shows the theoretical result for an angular-momentum conserving fluid.
The inset shows the corresponding radial monomer number density distributions for $f=10$ (full line) and $f=5$ (dashed line).
}
\end{figure}

\subsection{Ideal Star Polymers}\label{sec:stpol}
MPC simulations have been used intensively to investigate the behavior of macromolecules 
under flow~\cite{yeom06,male00b,ripo04,ripo06,webs05,lee06,wata07}.
Here, we consider a two-dimensional ideal star polymer with $f$ arms and arm length $L_f$ in a MPC fluid,
where the central monomer is fixed in the center of the enclosing cylinder with $R_2=10a$, 
which rotates with constant angular velocity $\Omega_0$.
Consecutive monomers are connected by the harmonic potential
$W_n=\frac{\kappa}{2} (\mathbf{r}_n-\mathbf{r}_{n+1})^2$ but are otherwise not interacting with each other.
The coupling to the fluid is achieved by including the monomers of mass $M$ 
in the collision step \cite{male00b}.
We simulate stars with $f=5$ and $f=10$ arms, both with an arm length $L_f=10$. We choose $M=5m_0$
and $\kappa=2 k_{\rm B}T/a^2$ for the spring constant, \ie $\langle (\mathbf{r}_n-\mathbf{r}_{n+1})^2 \rangle/a^2=1$ in equilibrium.

We determine the average azimuthal velocity of fluid particles and monomers as a function of $r$, 
as shown in Fig.~\ref{fig:star}.
While the MPC-AT$+a$ yields the physically correct result, we find a non-uniform angular velocity
in MPC-AT$-a$ fluid, similar to the case of the binary fluid, but without a sharp interface.
The star polymer, which is located at small radii (see density distribution 
in the inset of Fig.~\ref{fig:star}),
rotates more slowly than the cylinder with an average angular velocity
$\Omega_{\rm star}/\Omega_0=0.63 \pm 0.01$ for $f=10$ and 
$\Omega_{\rm star}/\Omega_0=0.74 \pm 0.01$ for $f=5$.
Note that for the chosen parameters, the radial monomer density near the center is quite large, 
see the inset of Fig.~\ref{fig:star}.
The effect of a reduced angular velocity is less pronounced for less compact stars, \ie 
for reduced arm number or decreased spring constant.
Thus, this artifact can be drastically reduced by keeping the local monomer density low,
for example by
taking into account excluded volume interactions.
The $-a$ methods should not be employed for high local density of embedded objects.

\section{Eccentric Cylinders \label{sec:ecc}}
\begin{figure}[ht]
\begin{center}
\includegraphics[angle=0,width=6cm,clip]{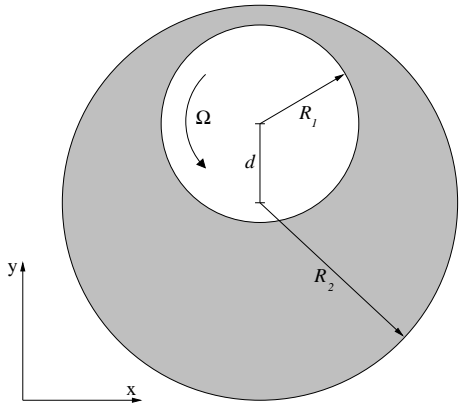}
\end{center}
\caption{Geometry a fluid (gray) enclosed between eccentric cylinders.}
\label{eccentric_geometry}
\end{figure}

\begin{figure}[ht]
\includegraphics{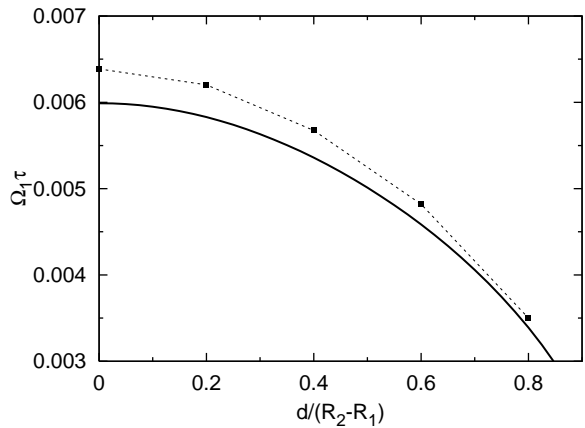}
\caption{Angular velocity $\Omega_1$ as a function of the axis offset $d$ 
for a constant torque $T=75 m_0\tau^{-2}a^2$. 
The squares are the results for the MPC-AT$+a$ simulation compared to 
the theoretical result of Ref. \cite{muel42} (full line), 
the dashed line serves as a guide to the eye. The radii of the cylinders are $R_1=10a$ and $R_2=20a$. 
The parameters for the fluid are $n=10$ and $\Delta t=0.05\tau$. 
Error bars are smaller than the symbol size and are therefore omitted.}
\label{eccentric_omega}
\end{figure}

\begin{figure}[ht]
\includegraphics{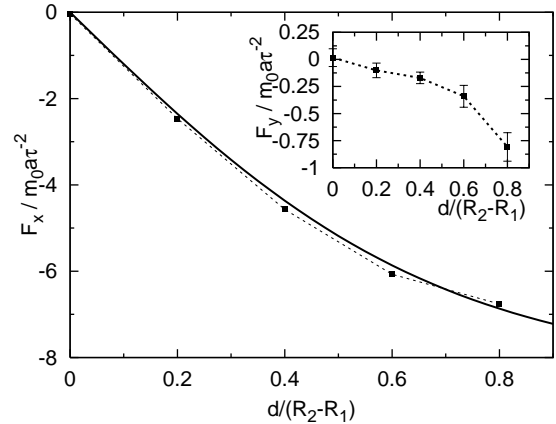}
\caption{Forces $F_x$, $F_y$ on the rotating inner cylinder perpendicular
and parallel to the line connecting the centers of the cylinders. The squares are simulation data obtained
from the MPC-AT$+a$ simulations, compared to the theory of Ref. \cite{muel42} (full line).}
\label{eccentric_force}
\end{figure}

Going one step further, we study a fluid between eccentric cylinders with radii $R_1$ and $R_2$ and fixed axes. 
The outer cylinder is stationary and the inner one is rotating about its axis 
with an angular velocity $\Omega_1$ (see Fig.~\ref{eccentric_geometry}).
Neglecting inertial forces, M\"uller \cite{muel42} derived theoretical expressions for the arising torques and 
forces acting on the cylinders, where the latter is predicted to be perpendicular to the line connecting the two centers.
We perform MPC-AT$+a$ simulations with constant torque, and measure the resulting angular 
velocities and the forces acting on the inner cylinder as a function of the axis offset $d$.
To avoid any bias, we use the simple 'cen'-boundary condition (see Sec. \ref{sec:boundary}).
The results are compared in Figs. \ref{eccentric_omega} and \ref{eccentric_force} 
with the theoretical predictions of Ref. \cite{muel42}.
In general, good agreement is found, 
although the cylinder is rotating up to 7\% faster than theoretically expected.
This can be explained by the finite slip on the surface of the cylinder; 
in order to suppress the slip completely, an extrapolation of the velocity field would be necessary for the virtual particles,
as discussed in Sec. \ref{sec:boundary}.
Moreover, we also observe a small radial component of the force, shown in the inset of Fig.~\ref{eccentric_force}.
It tends to move the inner cylinder to the center of the outer one, 
as it is expected when inertial effects are taken into account \cite{ball76}.

In the following we consider one specific geometry in more detail, depicted 
in Fig.~\ref{eccentric_geometry} with $R_1=10 a$, $R_2=20 a$ and an axis offset $d=8 a$, using the MPC-AT$+a$ algorithm.

\begin{figure}[ht]
\includegraphics{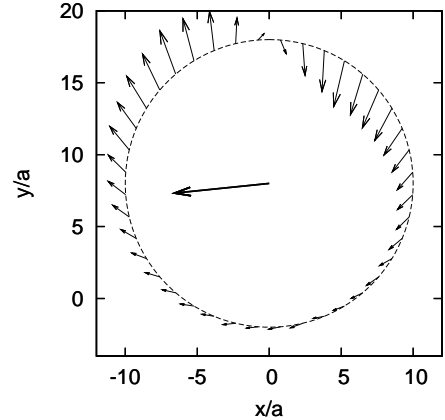}
\caption{Force distribution on the surface of the counter-clockwise rotating inner cylinder 
(indicated by the dashed line) for the geometry of Fig.~\ref{eccentric_geometry} (axis offset $d=8a$). 
A constant torque $T=75 m_0\tau^{-2}a^2$ is applied to the inner cylinder, resulting in a 
counter-clockwise rotation with $\langle \Omega_1 \rangle =0.0035\tau^{-1}$.
The length of the arrows is proportional to the local force. 
The direction of the resulting total force is marked by the thick arrow acting on the center of the cylinder.}
\label{eccentric_force_distribution}
\end{figure}

\begin{figure}[ht]
\includegraphics[angle=0,width=8cm,clip]{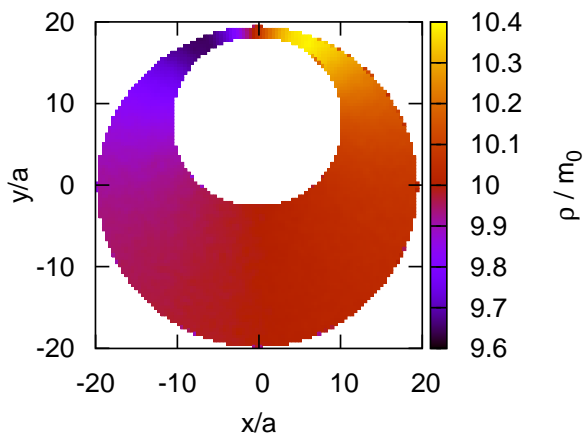}
\caption{(Color online) Density distribution of a MPC-AT$+a$ fluid enclosed by eccentric cylinders with $n=10$ 
and the same parameters as in Fig.~\ref{eccentric_force_distribution}.}
\label{eccentric_density}
\end{figure}

\begin{figure}[ht]
\begin{center}
\includegraphics[angle=0,width=7cm,clip]{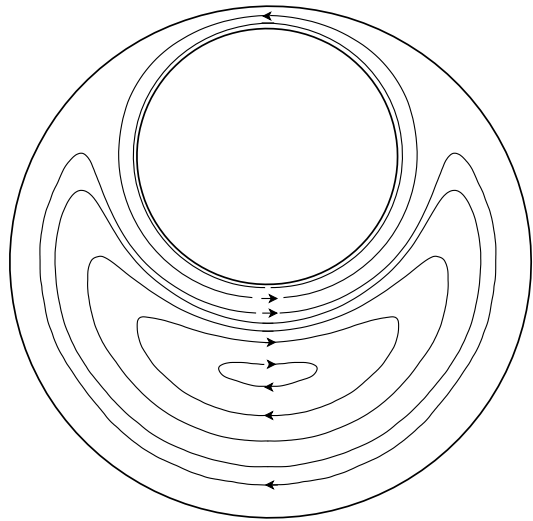}
\end{center}
\caption{Stream lines for the same parameters as in Fig.~\ref{eccentric_force_distribution}.}
\label{eccentric_streamlines}
\end{figure}

In Fig.~\ref{eccentric_force_distribution}, we present the measured force distribution 
on the surface of the inner cylinder, where the force due to the isotropic hydrostatic pressure has been subtracted.
The force exerted by the fluid on the cylinder is composed of two contributions:
First of all, the shear stress is hindering the counter-clockwise rotation.
This force tangential to the surface is more pronounced in the small-gap region 
(top of Fig.~\ref{eccentric_geometry} and \ref{eccentric_force_distribution}) than in the large-gap region, 
thus the net force due to viscous stress points in the positive $x$ direction, 
\ie to the right in Fig.~\ref{eccentric_geometry}.
Second, where the fluid is moving into and out of the slit, the dynamic pressure gives rise 
to regions of increased pressure on the right side and decreased pressure on the left (see Fig.~\ref{eccentric_density}).
This in turn induces a force pointing in the negative $x$ direction, 
counteracting the force due to viscous stress and exceeding the latter in strength, hence the total force points to the left.
Since the MPC fluid is compressible, density inhomogeneities emerge, but they 
are sufficiently small that the corresponding local variation of the viscosity is negligible.
The density distribution, which is proportional to the pressure distribution, 
is shown in Fig.~\ref{eccentric_density}.
The corresponding stream lines are shown in Fig.~\ref{eccentric_streamlines}, 
resembling very much the theoretical results of Refs.~\cite{ball76,maur97}.
In particular, a back-flow occurs in the large-gap region.

Next, we study how the results are affected by the lack of angular-momentum conservation.
Clearly, for a given external torque, in the MPC-AT$-a$ method the inner cylinder would rotate with an incorrect angular velocity.
As this effect has already been discussed in the preceding section, 
we fix the velocity on the boundary instead of imposing a constant torque.
We choose the results for the angular velocity obtained from the MPC-AT$+a$ simulations 
as an input parameter for the MPC-AT$-a$ simulations in order to investigate 
the influence of the angular-momentum conservation on the resulting velocity field.
For comparability, we chose the parameters in such a way that the viscosity $\eta=\bar{\eta}+\check{\eta}$ is the same 
for both simulation methods. We find practically identical velocity fields.
In order to quantify the difference, we calculate the ratio 
$\langle({\bf v}^{\rm AT+a}-{\bf v}^{\rm SR})^2\rangle^{1/2}/\langle({\bf v}^{\rm AT+a})^2\rangle^{1/2} \approx 0.03$, 
where ${\bf v}^{\rm AT+a}$ and ${\bf v}^{\rm SR}$ denote the velocity fields obtained 
by the two different simulation methods, and the average is taken over the simulation box.

Although the density distributions for both simulation methods are qualitatively very similar, 
the density inhomogeneity is slightly less pronounced for MPC-AT$-a$, giving rise to a smaller pressure gradient.
This is also reflected in the smaller total force acting on the rotating inner cylinder in the MPC-AT$-a$ simulation:
For the MPC-AT$-a$ method we find $F_x=(-6.16 \pm 0.011)m_0a\tau^{-2}$ 
compared to $F_x=(-6.76 \pm 0.014)m_0a\tau^{-2}$ in the MPC-AT$+a$ simulation.

\section{Summary \label{sec:sum}}

We have investigated the relevance of angular-momentum conservation in mesoscale hydrodynamics simulations.
We have focused on MPC methods, but similar results are also expected in other hydrodynamic methods 
without angular-momentum conservation, such as DPD$-a$~\cite{nogu07}.
Focusing on fluids confined between rotating cylinders, we compare two simulation variants 
that only differ in the conservation of angular momentum.

In the bulk, both simulation methods show physically correct flow behavior.
Here, the negligence of angular-momentum conservation simply leads to a modified viscosity.
However, we find that in situations where torques are acting on surfaces, often quantitative or even qualitative 
incorrect results are obtained without angular-momentum conservation.
In particular, there are non-physical torques occurring even in the rigid body rotation where no torque is expected.
This can be well understood from basic continuum fluid mechanics as the non-conservation of angular momentum 
gives rise to an asymmetric stress tensor.

The angular-momentum conservation is essential to be taken into account 
in the following cases to avoid non-physical torques.
(i) The boundary condition on walls is given by forces including torques,
such as in circular Couette flow.
(ii) Finite-sized objects with angular degrees of freedom,
or densely distributed point-like objects,
rotate in fluids by the hydrodynamic stress, such as in colloidal and polymer suspensions.
(iii) Fluids with different viscosities are in contact.
When the boundary conditions are given by velocities, $-a$ methods give the correct velocity field.
For example, MPC-SR reproduces the frequency of von Karman vortex shedding observed in experiments and other numerical methods 
(see the Strouhal number in Fig. 5 of Ref.~\cite{lamu01}).
Thus, the $+a$ version of mesoscale hydrodynamics methods have to be employed whenever torques play a role in the flow of (complex) fluids.

\begin{acknowledgments}
We would like to thank A.~Lamura (CNR Bari) and C. Bechinger (Stuttgart) for stimulating discussions. 
We acknowledge support of this work by the DFG 
through the priority program ``Nano- and Microfluidics''.
\end{acknowledgments}

\begin{appendix}

 \begin{center}
      {\bf APPENDIX: Calculation of Torque on Cylinder Surfaces}
    \end{center}

In order to calculate the momentum crossing a cylinder of radius $R$, we use the following simplification:
Instead of explicitly taking into account all quadratic cells that are intersected by $R$, 
due to the cylindrical symmetry of the problem 
it is favorable to consider an annular arc with radial width $a$ and area $a\theta_{\rm c}R_{\rm c}=a^2$ as an 'adapted' collision cell,
where $\theta_{\rm c}$ is the angular width and
the inner or outer radius is $R_{\rm c}-a/2$ or $R_{\rm c}+a/2$, respectively. 
The collision step locally equalizes the velocity within the cell on average, 
hence the pre-collisional velocity distribution $\mathbf{v}(\mathbf{r})=\Omega r {\bf e}_{\theta}$ 
is converted into the average azimuthal velocity
\begin{eqnarray}
v^{\rm c}_\theta(R_{\rm c}) &=&  \int \limits_{R_{\rm c}-a/2}^{R_{\rm c}+a/2} dr \int \limits_{-\theta_{\rm c}/2}^{\theta_{\rm c}/2} d\theta\  \frac{\Omega r^2 \cos(\theta)}{a^2} \nonumber \\ 
&\simeq& \Omega R_{\rm c}\left( 1+ \frac{a^2}{24{R_{\rm c}}^2} \right).
\end{eqnarray}
This collision accelerates or decelerates the fluid particles inside or outside of 
a virtual cylinder with radius $R$ in fluids, respectively.
We now calculate the change of angular momentum 
$\Delta L_{{\rm in},{\rm out}}(R,R_{\rm c})=L_{{\rm in},{\rm out}}^\prime(R,R_{\rm c})-L_{{\rm in},{\rm out}}(R,R_{\rm c})$ caused by this alteration 
of the velocity distribution,
where the subscript 'in' or 'out' denotes the inner or outer surfaces of a cylinder of radius $R$ and
 $L_{{\rm in},{\rm out}}$ or $L^\prime_{{\rm in},{\rm out}}$ denotes the angular momenta before or after the collision step, respectively.
For the inner sub-annulus, we find
\begin{equation}
\Delta L_{\rm in}(R,R_{\rm c}) = \int \limits_{R_{\rm c}-a/2}^R dr \  \frac{2\pi m(n-1)r^2}{a^2} [v^{\rm c}_\theta(R_{\rm c}) - v_\theta(r)]
\end{equation}
and analogously for the outer sub-annulus
\begin{equation}
\Delta L_{\rm out}(R,R_{\rm c}) = \int \limits_R^{R_{\rm c}+a/2} dr \  \frac{2\pi m(n-1)r^2}{a^2} [v^{\rm c}_\theta(R_{\rm c}) - v_\theta(r)].
\end{equation}
In the derivation, $\langle {\bf v}_i-{\bf v}_{\rm c}^{\rm G} \rangle=(1-1/n)(\langle {\bf v}_i \rangle -v^{\rm c}_\theta(R_{\rm c}){\bf e}_{\theta})$ is employed.
Note that $\Delta L_{\rm in}(R,R_{\rm c})$ and $\Delta L_{\rm out}(R,R_{\rm c})$ are not exactly oppositely equal, 
reflecting the fact that the total angular momentum of the considered annuli slightly changes.

To take into account the random grid shift, 
we subsequently average over all annuli containing $R$, \ie $R-a/2 \le R_{\rm c} \le R+a/2$.
Finally, the torque $T_{{\rm in},{\rm out}}= \langle\Delta L_{{\rm in},{\rm out}}(R,R_{\rm c})\rangle/\Delta t$ is given by
\begin{eqnarray}
T_{{\rm in},{\rm out}}(R) &=& \frac{\pi m(n-1) \Omega}{\Delta t} \Bigg[\pm \frac{5R^2}{36} - \frac{aR}{8} \pm \frac{29a^2}{1080} \nonumber \\
&& +\frac{1}{36} \bigg( \frac{a^2}{8} \pm \frac{R^3}{a} \bigg) \ln \frac{2R+a}{2R-a} \Bigg] \nonumber \\
&\simeq& \frac{\pi m(n-1) \Omega R^2}{6 \Delta t} \bigg(\pm 1 - \frac{3a}{4R} \bigg) \nonumber \\
&=&  4 \pi \check{\eta} \Omega R^2 \bigg(\pm 1 - \frac{3a}{4R} \bigg). 
\end{eqnarray} 

\end{appendix}

\bibliographystyle{apsrev}

\begin{thebibliography}{41}
\bibitem[{foo()}]{footnote1}
\bibinfo{note}{Alternatively, the new positions of the particles can be
  obtained by streaming them according to their velocities with reflections on
  both boundaries of the wall layers.}
\end{thebibliography}

\end{document}